\documentstyle[12pt,aps,prd]{revtex}
\newcommand{\beq}{\begin{equation}}
\newcommand{\eeq}{\end{equation}}

\def\lap{\lower.5ex\hbox{$\; \buildrel < \over \sim \;$}}
\def\gap{\lower.5ex\hbox{$\; \buildrel > \over \sim \;$}}
\def\L{\Lambda}
\def\rL{\rho_\Lambda}

\input epsf
\begin{document}

\title{On likely values of the cosmological constant}
\author{Jaume Garriga\/$^{1,2}$
and Alexander Vilenkin\/$^2$}
\address{
$^1$ IFAE, Departament de Fisica, Universitat Autonoma de Barcelona,\\
08193 Bellaterra (Barcelona), Spain\\
$^2$ Institute of Cosmology, Department of Physics and Astronomy,\\
Tufts University, Medford, MA 02155, USA}

\maketitle

\begin{abstract}

We discuss models in which the smallness of the effective vacuum
energy density $\rho_\L$ and the coincidence of the time of its
dominance $t_\L$ with the epoch of galaxy formation $t_G$ are
due to anthropic selection effects.  In such models, the
probability distribution for $\rho_\L$ is a 
product of an {\it a priori} distribution ${\cal P}_*(\rho_\L)$ 
and of the number density of galaxies at a given $\rho_\L$ (which
is proportional to the number of observers who will detect that value
of $\rho_\L$).  To determine ${\cal P}_*$, we consider
inflationary models in which the role of the vacuum energy is played
by a slowly-varying potential of some scalar field.  We show that the 
resulting distribution depends on the shape of the potential and
generally has a non-trivial dependence on $\rho_\L$, even in the
narrow anthropically allowed range.  This is contrary to Weinberg's
earlier conjecture that the {\it a priori} distribution should be nearly
flat in the range of interest.  We calculate the (final) probability
distributions for $\rho_\L$ and for $t_G/t_\L$ in simple
models with power-law potentials.  For some of these models, the
agreement with the observationally suggested values of $\rho_\L$
is better than with a flat {\it a priori} distribution. 
We also discuss quantum-cosmological approach in which $\rho_\L$
takes different values in different disconnected universes and argue
that Weinberg's conjecture is not valid in this case as well.
Finally, we extend our analysis to models of quintessence, with
similar conclusions.

\end{abstract}

\section{Introduction}

The cosmological constant $\Lambda$ presents us with a number of
perplexing problems (see \cite{sast99} for a recent review). 
Particle physics models suggest that the natural
value for $\Lambda$ is set by the Planck scale, $m_{pl}\sim 10^{19}$
GeV.  The corresponding vacuum energy density is 
\beq
\rho_\Lambda\sim m_{pl}^4,
\label{rhopl}
\eeq
which is some 120 orders of magnitude greater than the observational
bounds.  This is what is usually called ``the cosmological constant
problem''.  The discrepancy between the expected and observed values
is so large that until recently it was almost universally believed
that the cosmological constant must vanish.  However, no convincing
mechanism has yet been found that would set $\L$ to zero.  

It came as a total surprise when recent observations \cite{Supernova} 
provided
strong evidence that the universe is accelerating, rather than
decelerating, suggesting a non-zero cosmological constant.  While
there was still hope to explain a vanishing $\Lambda$, a small
non-zero value appeared totally incomprehensible.  

The observationally suggested values of $\L$ correspond to
$\rho_\L\sim\rho_{M0}$, where $\rho_{M0}$ is the present density of
matter.  This brings yet another puzzle.  It is difficult to
understand why we happen to live at the epoch when
$\rho_M\sim\rho_\L$.  That is, why
\beq
t_0\sim t_\L,
\label{t0tL}
\eeq
where $t_0$ is the present time and $t_\L$ is the time when the
cosmological constant starts to dominate.  Observers living at $t\ll
t_\L$ would find $\rho_M\gg\rho_\L$, while observers living at $t\gg
t_\L$ would find $\rho_M\ll \rho_\L$.

The only explanation of these puzzles that we are aware of attributes
them to anthropic selection effects.  In this approach, the
cosmological constant is assumed to be a free parameter that can take
different values in defferent parts of the universe, or perhaps in
different disconnected universes.  Weinberg \cite{Weinberg87} was the
first to point out that not all values of $\L$ are consistent with the
existence of conscious observers \cite{Davis}.  
In a spatially flat universe with a
cosmological constant, gravitational clustering effectively stops at
$t\sim t_\L$, corresponding to the
redshift $(1+z_\L)\sim (\rho_\L/\rho_{M0})^{1/3}$.  At later times,
the vacuum energy dominates and the universe enters a de Sitter stage
of exponential expansion.  An anthropic bound on $\rho_\L$ can be
obtained by requiring that it does not dominate before the redshift
$z_{max}$ when the earliest galaxies are formed,
\beq
\rho_\L\lesssim (1+z_{max})^3 \rho_{M0}.
\eeq
Weinberg took $z_{max}\sim 4$, which gives 
\beq
\rho_\L\lesssim 100\rho_{M0}.  
\label{Wbound}
\eeq
This is a dramatic improvement over Eq.(\ref{rhopl}), but it still
falls short of the observational bound by a factor of about 30.

The anthropic bound (\ref{Wbound}) specifies the value of $\rho_\L$
which makes galaxy formation barely possible.  However, as it was
pointed out in \cite{AV95,Efstathiou}, 
the observers are where the galaxies are, and
thus most of the observers will detect not these marginal values, but
rather the values that maximize the number of galaxies.  More
precisely, the probability distribution for $\rho_\L$ can be written
as \cite{AV95}
\beq
d{\cal P}(\rL)={\cal P}_*(\rL)\nu(\rL)d\rL.
\label{dP}
\eeq
Here, ${\cal P}_*(\rL)d\rL$ is the {\it a priori} distribution, which
is proportional to the volume of those parts of the universe where
$\rho_\L$ takes values in the interval $d\rL$, and $\nu(\rL)$ is the
average number of galaxies that form per unit volume with a given
value of $\rL$.  The calculation of $\nu(\rL)$ is a standard
astrophysical problem; it can be done, for example, using the
Press-Schechter formalism \cite{PS}.  The {\it a priori} distribution 
${\cal P}_*(\rL)$ should be determined from the theory of initial
conditions, e.g., from an inflationary model or from quantum
cosmology.  

Martel, Shapiro and Weinberg \cite{MSW} (see also \cite{Weinberg96})
presented a detailed calculation of $d{\cal P}(\rL)$.
They first noted that ${\cal P}_*(\rL)$ can be
expected to vary on some characteristic particle physics scale,
$\Delta\rL\sim\eta^4$.  The energy scale $\eta$ could be anywhere
between the Planck scale and the electroweak scale, $\eta_{EW}\sim
10^2$ GeV.  For any reasonable choice of $\eta$, $\Delta\rL$ exceeds
the anthropically allowed range (\ref{Wbound}) by many orders of
magnitude. Also, in the absence of a mechanism that
sets the cosmological constant to zero, we may not expect any pronounced
features in the probability distribution at low values of $\rL$.
This suggests that we can set
\beq
{\cal P}_*(\rL)=const
\label{WC}
\eeq
in the range of interest.  This argument is originally due to Weinberg
\cite{Weinberg87}, and we shall refer to (\ref{WC}) as 
Weinberg's conjecture.
Once it is accepted, the problem reduces to the calculation of
$\nu(\rL)$.  Martel et.al. found that 
the resulting probability distribution is peaked at
somewhat larger values of $\rL$ than observationally suggested.  For
the probability of $\rL$ being smaller or equal than the values indicated 
by the supernova data, it gives $\sim 5-10\%$. In absolute terms, 
this is not a very large probability. However, the mere fact that it is
non-negligible is rather impressive, in view of the large 
discrepancy in orders of magnitude between the {\em a priori} expected 
range for $\rL$ and its measured value. 

Going back to the issue of the cosmic time coincidence,
(\ref{t0tL}), this can also be explained by
anthropic selection effects.  Here is a sketch of the argument
\cite{Foot0,GLV,Bludman}.  One first notes that the
present time $t_0$ is bounded by 
\beq
t_G\lesssim t_0\lesssim t_G +t_\star, 
\label{ts}
\eeq
where $t_G$ is the time of galaxy formation (which is also
the time when most of the stars are formed) and $t_\star$
is the characteristic lifetime of habitable stars, $t_\star\sim 5-20$
Gyr.  Observationally, giant galaxies were assembled at $z\sim 1-3$,
or $t_G\sim t_0/3-t_0/8$, that is, within an order of magnitude of
$t_0$.  Since $t_G\sim t_\star$, it follows from (\ref{ts}) that 
most observers live at the epoch when $t\sim t_G$, and
the problem of explaining the coincidence $t_0\sim t_\L$ is reduced to
explaining why
\beq
t_G\sim t_\L.
\label{tGtL}
\eeq
The latter coincidence is not difficult to understand if we note that
regions of the universe where $t_\L \ll t_G$ do not form any galaxies
at all.

The ``coincidence'' (\ref{tGtL}) can be expressed quantitatively by
calculating the probability distribution for $t_G/t_\L$.  With a flat
{\it a priori} distribution (\ref{WC}), one finds \cite{GLV} that it
has a broad peak in the range $0.3\lesssim t_G/t_\L\lesssim 5$ with a
median  at $t_G/t_\L\approx 1.5$.  Thus, most observers will find
themselves living in galaxies formed at $t_G\sim t_\L$.

The probability distributions for $\rL$ and $t_G/t_\L$ were
calculated in Refs. \cite{MSW,GLV} by using
Weinberg's conjecture. That is, without recourse to any particular
model that would allow $\Lambda$ to vary and simply assuming the flat
distribution (\ref{WC}). This is the beauty of the conjecture:
if true, it would make the results independent of one's (necessarily
speculative) assumptions about the very early universe, and therefore
it would make the theory more predictive.  It is important, however, to
consider specific models with a variable vacuum energy and to check
whether or not the conjecture is actually valid.  This is one of our
goals in the present paper.

In the next section we discuss models in which the role of the
cosmological constant is played by a very slowly varying potential
$V(\phi)$ of some scalar field $\phi$.  We find that, unfortunately,
Weinberg's conjecture is not generally valid in such models, and that
the {\it a priori} distribution ${\cal P}_*(\rho_\phi)$ can be
expected to be a non-trivial function of $\rho_\phi$ in the range of
interest [here, $\rho_\phi\equiv V(\phi)$].  We give some examples of
potentials which do and do not satisfy the conjecture.

In Section III, we use simple models with power-law potentials,
$V(\phi)\propto\phi^n$,  to study
the effect of a non-trivial {\it a priori} distribution on the final
probability distribution for $\rho_\phi$ and on the cosmic time coincidence.
For some of these models, we find that the agreement with
the observationally suggested values of $\rho_\phi$ is better than
what one gets from the calculations based on the flat distribution
(\ref{WC}). 

In Section IV we discuss models in which $\Lambda$ does not change
throughout the universe but may take a range of values in different
disconnected universes.  Once again, we argue that Weinberg's
conjecture is not likely to be valid in this case.

In Section V we extend our approach to models of quintessence.  
Our conclusions are briefly summarized in Section VI.

\section{Slowly varying potentials}

Suppose that what we
perceive as a cosmological constant is in fact a potential $V(\phi)$ of
some field $\phi(x)$.  Observations will not distinguish between
$V(\phi)$ and a true cosmological constant, provided that the kinetic
energy of $\phi$ is small compared to $V(\phi)$,
\beq
{\dot\phi}^2/2\ll V(\phi).
\label{kinpot}
\eeq
The evolution of $\phi$ is then described by the slow roll equation
\beq
3H{\dot\phi}\approx -V'(\phi),
\label{slowroll}
\eeq
and Eq.(\ref{kinpot}) gives
\beq
{V'}^2\ll 18 H^2 V.
\label{C}
\eeq
We want to require that this condition still applies at the time when
$V(\phi)$ is about to dominate.  Then 
\beq
H^2\sim 8\pi V(\phi)/3 m_{pl}^2, 
\label{H}
\eeq
and Eq.(\ref{C}) yields
\beq
|V'(\phi)|\ll 12 V(\phi)/m_{pl}.
\label{flatpot}
\eeq

The dynamics of the field $\phi$
during inflation are strongly influenced by quantum fluctuations, causing
different regions of the universe to thermalize with different values of
$\phi$.  Spatial variation of $\phi$ is thus a natural outcome of
inflation.  If $V(\phi)$ is sufficiently small, its back reaction on
the rate of inflationary expansion is negligible, and all values of
$\phi$ are equally probable,
\beq
d{\cal P}_*(\phi)\propto d\phi.
\label{flatphi}
\eeq 
The condition for negligible back reaction is \cite{Foot1}
\beq
m_{pl}^2 {V'}^2 H^4/V^3\gg 1,
\label{back}
\eeq
where here $H$ is the Hubble rate during inflation.

Let us now recall
that Weinberg's conjecture was motivated by the fact that the
anthropically allowed range of $\rho_\phi$ is very small compared to
the natural range of variation of $\rho_\phi$.  One could expect that
in this small range $V(\phi)$ can be approximated by a linear
function.  With an appropriate choice for the origin of $\phi$,  
\beq
V(\phi)=\kappa \phi. 
\label{linear}
\eeq
Then Eq.(\ref{flatphi}) implies a flat distribution for the vacuum energy
density $\rho_\phi\equiv V(\phi)$,
\beq
d{\cal P}_*(\rho_\phi)\propto d\rho_\phi.
\label{flat'}
\eeq
However, in our case Eq.(\ref{flatpot}) 
applied to the present time requires that 
the slope of $V(\phi)$ should be extremely small. The present Hubble rate is 
$H_0\sim 10^{-61} m_{pl}$, so using (\ref{flatpot}) and (\ref{H}) we have 
$\kappa \lesssim 10^{-122} m_{pl}^3$.
As a result, a small range of $V(\phi)$ may correspond to a very large
range of $\phi$.
Indeed, it follows from (\ref{flatpot}) that 
\beq
\Delta\phi\sim V/V'\gg m_{pl}.
\label{deltaphi}
\eeq
The natural range of $\phi$ in particle physics models is
$\Delta\phi\lesssim m_{pl}$, and there seems to be no reason to expect
the slope of $V(\phi)$ to remain constant over the super-Planckian
range (\ref{deltaphi}).

Thus, we conclude that (i) models with variable $\rho_\phi$ can be
easily constructed in the framework of inflationary cosmology
and that
(ii) Weinberg's conjecture (\ref{WC}) will not generally apply in this
class of models.  In the general case,
assuming negligible
back-reaction, Eq.(\ref{flatphi}) yields 
\beq
d{\cal P}_*(\rho_\phi)\propto [V'(\phi)]^{-1}d\rho_\phi.
\label{PV}
\eeq
We now discuss some examples of potentials that do and do not satisfy
the conjecture.

\subsection{Some examples}


We first consider a scalar field with a quadratic potential,
\begin{equation} 
V(\phi)= \rL+ {\mu^2 \over 2} \phi^2,
\label{quad}
\end{equation}
where $\rL$ is a ``true'' cosmological constant, which is assumed to be
large.  We assume also that $\rL$ and $\mu^2$ have opposite signs, so
that the two terms in (\ref{quad}) partially cancel one another
in some parts of the universe.  

The cancellation occurs for $\phi\sim
\sqrt{\rL}/\mu$, and 
Eqs. (\ref{flatpot}), (\ref{H}) give the condition $\mu\ll H_0^2
m_p/\sqrt{\rL}$, where $H_0\sim 10^{-61}m_{pl}$ is the present Hubble
rate.  With $\rL\sim m_p^4$, this gives 
\begin{equation}
|\mu|\ll 10^{-122} m_{pl}.
\label{smallmass}
\end{equation}
Thus, an exceedingly small mass scale must be introduced.
On the other hand, the condition (\ref{back}) for negligible backreaction
imposes 
\beq
|\mu|\gg H^3_0 H^{-2} \sim 10^{-169} m_{pl},
\eeq 
where we have used $H\sim 10^{-7} m_{pl}$, corresponding to a GUT
scale of inflation.

A critical reader may wonder at this point if anything is going to be
achieved by explaining a cosmological constant $\rho_\L\sim
10^{-120}m_{pl}^4$ 
in terms of a scalar field with a 
small mass of order $|\mu|\ll 10^{-122} m_{pl}$.
However, potentials with very small masses or couplings
could be generated through instanton effects.  Suppose that we
have a field $\phi$ with a flat potential, $V(\phi)=const.$, and that
the radiative corrections to $V(\phi)$ vanish to all orders of
perturbation theory, due to some symmetry.  (For example, $\phi$ could
be a Goldstone boson.)  Suppose further that the symmetry is violated
by instanton effects.  Then $\phi$ will acquire a mass of the order
$\mu^2 \sim e^{-S} m_{pl}^2$, 
where $S$ is the instanton action.  In order to
have $|\mu| \ll 10^{-122} m_{pl}$, one needs $S \gtrsim 560$, which is 
not unreasonable.  

The critic may still be unsatisfied and ask why the
same kind of argument cannot be applied directly to the cosmological
constant.  One could imagine that $\rL=0$ to all orders of
perturbation theory, due to some approximate symmetry, 
and that a small $\rL\propto \exp (-S)$ is induced by instantons.  The
problem with this scenario is that it does not explain the cosmic time
coincidence (\ref{t0tL}).  The instanton action $S$ should be
fine-tuned so that $\L$ starts dominating at the present time.  
Models with $\rL$ replaced by $V(\phi)$ are therefore preferred.

The potential (\ref{quad}) can be rewritten  as 
\beq
V(\phi)\equiv \rho_\phi=
\kappa (\phi-\phi_0) + {\mu^2\over 2} (\phi-\phi_0)^2,
\eeq
where $\phi_0^2=-2\rL/\mu^2$ and $\kappa=\mu^2\phi_0$.
Then, using (\ref{PV}) in the vicinity of $\phi=\phi_0$ we have
\beq
d{\cal P}_*(\rho_{\phi})
\propto  \left(1+ 2{ \mu^2\over \kappa^2}\rho_{\phi}\right)^{-1/2} 
d\rho_{\phi} = [1+{\cal O}(\rho_{\phi}/\rL)]\ d\rho_{\phi}.
\eeq
Since $\rho_{\phi}/\rL \ll 1$ in the anthropically allowed range, we conclude
that Weinberg's conjecture applies to very good approximation in this
case.  The reason is that the cancellation of the two terms in
(\ref{quad}) occurs at a very large value of $\phi\gg m_{pl}$ and the
characteristic range of variation of a power-law potential is
$\Delta\phi\sim\phi$.  For the same reason, potentials of the form
\beq
V(\phi)=\rL+A\phi^n
\eeq
can also be expected to satisfy the conjecture.

To give an example of a potential for which Weinberg's conjecture is
not satisfied, consider a ``washboard'' potential
\beq
V(\phi)=\rL+A\phi +B\sin(\phi/M),
\eeq
where $M\lesssim m_{pl}$ is some particle physics scale and the
constants $A$ and $B$ are small enough to satisfy the condition
(\ref{flatpot}).   In this case, Eq.(\ref{PV}) gives a distribution
\beq
{\cal P}_*\propto [A+(B/M)\cos (\phi/M)]^{-1}, 
\eeq
which is not flat, unless $B/AM\ll 1$.

\section{Power-law potentials}

We shall now consider a different situation,
where the true cosmological constant has been set equal to 
zero by some unspecified mechanism, but the potential energy of a scalar
field (whose minimum is at $V=0$) induces a small effective 
cosmological constant.  Since the minimum of the potential is at
$\rho_\phi=0$, Weinberg's conjecture is not expected to apply in
this case.  
  
To illustrate the effects of a non-trivial {\it a priori} distribution
${\cal P}_*(\rho_\phi)$,
we shall calculate the probability distributions
for $\rho_\phi$ and $t_G/t_\phi$ in the simple case of a power-law potential,
\beq
V(\phi)\propto \phi^n.
\label{powerlaw}
\eeq
Familiar examples of such potentials are
\beq
V(\phi)={1\over{2}}m^2\phi^2
\label{Vm}
\eeq
and
\beq
V(\phi)={1\over{4}}\lambda\phi^4.
\label{Vlambda}
\eeq
They can be suitable for our purposes only if the parameters $m$ and
$\lambda$ are very small.  Indeed, Eqs.(\ref{H}) and (\ref{flatpot})
require $\phi\gg m_{pl}/6$, $m\ll 3H_0$ for (\ref{Vm}) and $\phi\gg
m_{pl}/3$, $\lambda\ll 40 H_0^2/m_{pl}^2$ for (\ref{Vlambda}).  Thus, we
obtain the constraints $m\ll 10^{-61}m_{pl}$ and $\lambda\ll
10^{-119}$. The condition (\ref{back}) for negligible backreaction
will impose lower bounds on these parameters. 
For the quadratic potential it requires
$m\gg 10^{-108}m_{pl}$ and for the quartic it gives $\lambda \gg 10^{-310}$.
Here, as in the previous section, we
have used $H\approx 10^{-7} m_{pl}$, corresponding to a GUT-scale
inflation,
and $V\sim m_{pl}^2 H_0^2$, with $H_0\approx 10^{-61} m_{pl}$.  
Inflation at a lower energy scale will impose somewhat tighter bounds.
Again, the small masses and couplings can be induced by instanton effects.

In what follows we shall assume that back-reaction effects are
negligible
(Otherwise, ${\cal P}_*(\phi)$ can be calculated by
solving the Fokker-Planck equation of stochastic inflation; see
Ref. \cite{VVW}).  Then, substituting (\ref{powerlaw}) in (\ref{PV}) we
have
\beq
d{\cal P}_*(\rho_\phi)\propto \rho_\phi^{-{{n-1}\over{n}}} d\rho_\phi.
\label{P*}
\eeq
For $n>1$, the probability density grows towards smaller values of
$\rho_\phi$ and has an integrable singularity at $\rho_\phi=0$.  For
$n=1$, the distribution is flat, as in Weinberg's conjecture
(\ref{WC}).  For $0<n<1$, it grows towards large values of
$\rho_\phi$.  Finally, for $n<0$ the distribution has a non-integrable
singularity at $\rho_\phi=0$; in this case $\rho_\phi=0$ with a 100\%
probability.  As we mentioned in the Introduction, for a flat
{\it a priori} distribution ${\cal P}_*(\rho_\phi)=const~$ ($n=1$), the
full probability distribution (\ref{dP}) is peaked at a somewhat
larger value of $\rho_\phi$ than observationally indicated.  The
agreement with observations may be improved if ${\cal P}_*(\rho_\phi)$
grows towards smaller values as for $n>1$.  We shall therefore
concentrate on this case.

Following \cite{GLV}, we introduce the variable 
\begin{equation}
x={\Omega_\phi\over{\Omega_M}}=\sinh^2\left({t\over t_{\phi}}\right),
\label{x}
\end{equation}
where $\Omega_M$ and $\Omega_\phi$ are, respectively, the densities of
matter and of the scalar potential in units of the critical density,
and $t_\phi$ is the time of $\phi$-domination.
For convenience, we have defined $t_{\phi}$ as the time at
which $\Omega_{\phi}= \sinh^2(1)\Omega_M \approx 1.38 \Omega_M$.
At the time of recombination, for values of $\rho_\phi$ within the
anthropic range, $x_{rec}\approx \rho_\phi/\rho_{rec}\ll 1$, where the
matter density at recombination, $\rho_{rec}$, is independent of
$\phi$.  We can therefore express the probability distribution for
$\rho_\phi$ as a distribution for $x_{rec}$,
\beq
d{\cal P}(x_{rec})\propto \nu(x_{rec}) x_{rec}^{1/n} d\ln x_{rec},
\label{Pxrec}
\eeq
where $\nu(x_{rec})$ is the number of galaxies formed per unit volume
in regions with a given value of $x_{rec}$. For $n=1$ the calculation of the
distribution (\ref{Pxrec}) was discussed in detail by Martel
et. al. \cite{MSW}. In \cite{GLV} we gave a simplified version of their 
calculation, which we generalize here to the case $n>1$.

In a universe where the effective cosmological constant is non-vanishing, 
a primordial overdensity will eventually collapse provided that
its value at the time of recombination exceeds a certain critical value
$\delta^{rec}_{c}$. In the spherical collapse model this is estimated as 
$\delta^{rec}_{c}=1.13\  x_{rec}^{1/3}$ (see e.g. \cite{masha}). 
Hence, the fraction of matter that eventually clusters in galaxies 
can be roughly approximated as \cite{PS,masha}:
\begin{equation}
\nu(x_{rec})\approx 
{\rm erfc}\left({\delta^{rec}_c\over \sqrt{2}\sigma_{rec}(M_g)}\right)
\approx 
{\rm erfc}\left({.80\ x_{rec}^{1/3}\over\sigma_{rec}(M_g)}\right).
\label{nu}
\end{equation}
Here, erfc is the complementary error function and
$\sigma_{rec}(M_g)$ is the dispersion in the density contrast at the
time of recombination on the relevant galactic mass scale $M_g \sim 10^{12}
M_{\odot}$. 

The logarithmic distribution $d{\cal P}/d\ln\ x_{rec}=x_{rec}^{1/n}
\nu(x_{rec})$ 
is plotted in Fig. 1 for several values of $n$.
For $n=1$ it has a rather broad peak which spans two orders of magnitude in 
$x_{rec}$, with a maximum at 
\begin{equation}
x^{peak}_{rec}\approx 2.45\ \sigma_{rec}^3.
\label{xpeak}
\end{equation}
As noted by Martel et al. \cite{MSW}, the parameter $\sigma_{rec}$ 
can be inferred 
from observations of the cosmic microwave background anisotropies, although
its value depends on the assumed value of the effective 
cosmological constant in our part of the universe today.  
For instance, assuming that the present cosmological constant is
$\Omega_{\phi,0}=.8$, and the relevant galactic co-moving scale is in 
the range $R= (1-2) Mpc$, Martel et al. found 
$\sigma_{rec} \approx (2.3-1.7)\ 10^{-3}$. In this estimate, they also 
assumed a
scale invariant spectrum of density perturbations, a value
of $70 km\ s^{-1}\ Mpc^{-1}$ for the present Hubble rate,  and they defined
recombination to be at redshift $z_{rec}\approx 1000$ (this definition is 
conventional, since the probability distribution for the cosmological 
constant does not depend on the choice of reference time). Thus, taking
into account that $x$ scales like $(1+z)^{-3}$ in equation (\ref{xpeak}),
one finds that the peak of the distribution for the cosmological constant
today is at $x_0^{peak} \approx 29.8 - 12$. The value corresponding
to the assumed $\Omega_{\phi,0}=.8$ is $x_0 = 4$, certainly within
the broad peak of the distribution and not far from its maximum.

\begin{figure}[t]
\centering
\hspace*{-4mm}
\leavevmode\epsfysize=10 cm \epsfbox{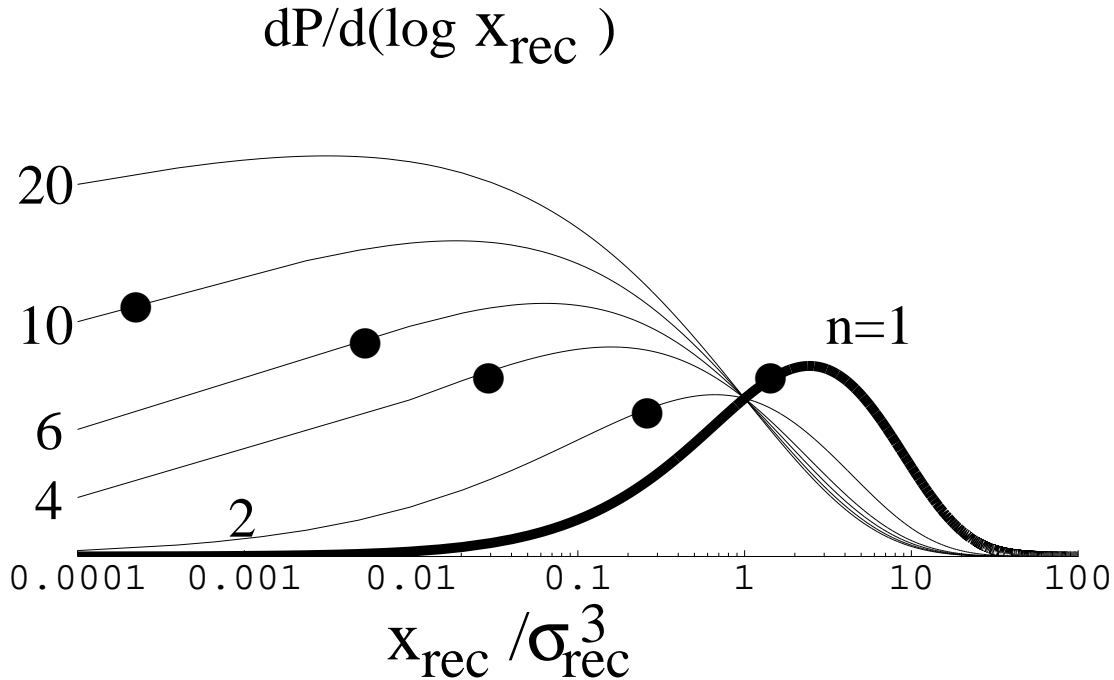}\\[3mm]
\caption[fig1]{\label{fig1} The probability distribution 
(\ref{Pxrec}) for the 
effective cosmological constant $\rho_{\phi}$, for different values of $n$.
As explained in the text, an observed value of $\Omega_\phi \approx .7$ 
corresponds to $x_{rec}/\sigma_{rec}^3 \approx .1$. There is at 
present some uncertainty in this estimate, because a number of assumptions 
must be made in order to infer the value of $\sigma_{rec}$ 
from observations. Notice, however, that this value lies at the tail of 
the $n=1$ curve, 
corresponding to Weinberg's conjecture (a flat {\em a priori} distribution). 
On the other hand, 
for $2\leq n \lesssim 5$ the value $x_{rec}/\sigma_{rec}^3 \approx .1$
is well within the broad peak
of the distribution. Thus, it is possible that a departure from Weinberg's 
conjecture may actually fit the observations better (more so
if it turns out that the cosmological constant is smaller than .7). 
The median of each distribution is indicated by a round bead.}
\end{figure}

However, if we assume instead that the measured value is 
$\Omega_{\phi,0}=.7$, which corresponds to $x_0= 2.3$, the new 
inferred values for $\sigma_{rec} \approx (3.3-2.4)\ 10^{-3}$
correspond to the peak value $x_0^{peak}\approx (88 - 34)$. In this case,
for $n=1$, the measured value would be at the outskirts of the broad peak, 
where the logarithmic probability density is about an order of magnitude 
smaller than at the peak. Although this is still a significant probability,
it is unfortunately somewhat low. 

For a potential (\ref{powerlaw}) with $n>1$, the peak in 
the distribution shifts to lower values of the effective cosmological 
constant, and therefore a measured value of $\Omega_{\phi,0}=.7$ 
(which corresponds to $x_{rec}\sigma_{rec}^{-3}\approx .1$) becomes much
better positioned. From Fig. 1, it is clear that for $2\leq n \lesssim 5$ this 
value lies well within the broad peak of the distribution. Thus we conclude
that the violation of Weinberg's conjecture by a power-law potential 
with $n>1$ may actually lead to a better agreement with observations.

\begin{figure}[t]
\centering
\hspace*{-4mm}
\leavevmode\epsfysize=10 cm \epsfbox{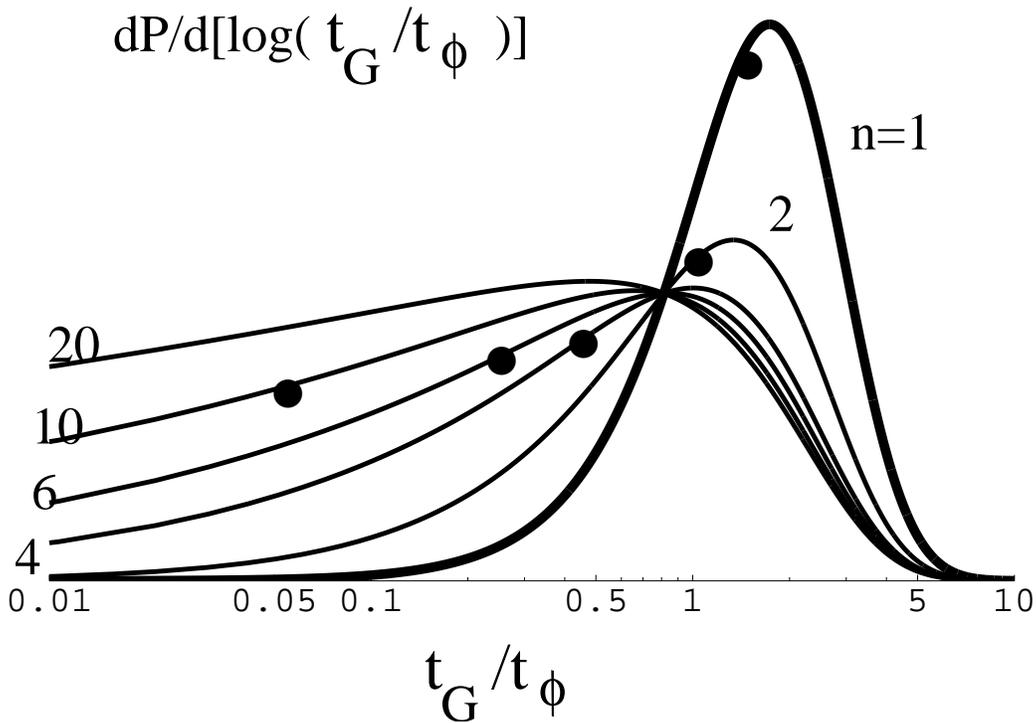}\\[3mm]
\caption[fig2]{\label{fig2} Probability distribution for $t_G/t_{\phi}$,
Eq. (\ref{prob3}), for different values of $n$. The round beads indicate
the median of each distribution. Note that the time
coincidence $t_G\sim t_\phi$} is not unexpected for $1\leq n \lesssim 10$.
\end{figure}

Let us now consider the issue of the time coincidence. Following our
earlier computation \cite{GLV} for the case $n=1$, we find that 
the probability distribution for $t_G/t_\phi$ is given by
\beq
d{\cal P}(t_G/t_{\phi}) 
\propto  [F(x)]^{{3\over n}-1} 
F'(x) {dx\over d\ln(t_G/t_{\phi})}
d\ln(t_G/t_\phi),
\label{prob3}
\eeq
where, here, $x=\sinh^2(t_G/t_\phi)$ and
\beq 
F(x) ={5\over 6} \left({1+x \over x}\right)^{1/2} 
\int_0^x {d\omega \over \omega^{1/6}(1+\omega)^{3/2}}.
\eeq
This distribution is shown in Fig. 2 for various values of $n$.  
For $n=1$ it has a broad peak which
almost vanishes outside of the range
$.1\lesssim (t_G/t_{\phi}) \lesssim 10$.  The maximum of the
distribution is at $t_G/t_{\phi}\approx 1.7$ and the median value is
at $t_G/t_\phi\approx 1.5$. 
Thus, most observers will find that their galaxies
formed at $t\sim t_\Lambda$, which explains the time coincidence
\beq
t_G\sim t_\Lambda.
\label{tgtlambda}
\eeq
As shown in Fig. 2, smaller values of $t_G/t_{\phi}$ become
more likely as we increase $n$. However,
values of $n\lesssim 10$ do not really spoil 
the coincidence (\ref{tgtlambda}), and even for $n$ as large as $30$, 
there is still a $5 \%$ probability for having $t_G/t_{\phi}\geq 1$.

\section{Quantum cosmology}

Let us now consider models with a true cosmological constant,
$\rL=const$, which takes the same value in the entire universe but may
have different values in other disconnected universes.  One example
\cite{Hawking} is
given by a four-index field $F_{\mu\nu\sigma\tau}$ whose value is
undetermined by the field equations, $\partial_\lambda
F_{\mu\nu\sigma\tau}=0$, and which gives a constant contribution to
the vacuum energy,
\beq
\rL=-(1/48)F_{\mu\nu\sigma\tau}F^{\mu\nu\sigma\tau}.  
\eeq
The {\it a priori} probability distribution for $\rL$ in this kind of
models can be found in the 
framework of quantum cosmology \cite{QC}.  
One should calculate the cosmological
wave function $\psi(\rL)$ which gives an amplitude for an inflationary
universe to
nucleate with a given value of $\rL$.  In the semiclassical
approximation, 
\beq
\psi\propto e^{\pm S/2},
\label{psi}
\eeq
where $S$ is the action of the corresponding instanton.  The upper
sign in (\ref{psi}) is for the tunneling wave function and the
lower sign for the Hartle-Hawking wave function.  This choice of sign
is a matter of some controversy \cite{Debate}, 
but it will not be important for our
discussion here.  The nucleation probability corresponding to
(\ref{psi}) is
\beq
{\cal P}_{nucl}(\rL)\propto \exp [\pm S(\rL)].
\label{Pnucl}
\eeq

The instanton in (\ref{Pnucl}) is a solution of Euclidean Einstein's
equations (possibly with quadratic and higher-order curvature
corrections) with a cosmological constant $\rL$ and a high-energy inflaton 
potential as a source. 
For small values of $\rL$, one can expect the instanton action to be
independent of $\rL$, $S(\rL)\approx const$.   
We note, however, that different universes in the ensemble described
by the wave function $\psi$ will generally have very different numbers
of galaxies and, therefore, of observers.  To take this into account,
one has to use Eq.(\ref{dP}) with
\beq
{\cal P}_*(\rL)\propto {\cal P}_{nucl}(\rL){\cal V}_*(\rL),
\label{P*'}
\eeq
where ${\cal V}_*(\rL)$ is the volume of the universe at the end of
inflation, when the vacuum energy is thermalized.  [The factor
$\nu(\rL)$ in Eq.(\ref{dP}) should then be understood as the number of
galaxies formed per unit thermalized volume.]

The right-hand side of Eq.(\ref{P*'}) would be well defined 
if inflation had a finite duration, 
so that ${\cal V}_*(\rL)<\infty$.  
It is well known, however, that inflation is generically eternal 
\cite{AV83,Linde86}: 
at any time
there are parts of the universe that are still inflating, and both
inflating and thermalized volumes grow exponentially with time.  In an
ensemble of eternally inflating universes, all volumes ${\cal V}_*$
become infinite in the limit $t\to\infty$, and Eq.(\ref{P*'}) becomes
meaningless \footnote{In fact this conclusion seems to apply
even if the inflaton potential does not drive eternal inflation. 
After a finite period of inflation the cosmological constant will 
eventually dominate, driving a de Sitter-like phase. Recycling events 
\cite{rec} that create new regions of the inflating
phase will then occur at a constant rate per unit spacetime volume,
making the total thermalized volume an exponentially growing function of time.}
.

It appears reasonable, in this case, to look not at the total volume
${\cal V}_*$ but at the rate of its growth (which generally depends on
$\rL$).  With a cutoff at time $t$,
\beq
{\cal V}_*(\rL,t)\propto \exp[\gamma(\rL)t],
\label{Vt}
\eeq
and the most probable value $\rL^{(*)}$ is found from
\cite{AV95,Linde95,Foot2}
\beq
\gamma(\rL^{(*)})=max.
\label{gammamax}
\eeq
As time goes on, the volume of the universes with this preferred value
of $\rL$ gets larger than the competition by an arbitrarily large
factor, and thus in the limit $t\to\infty$ the probability for
$\rL=\rL^{(*)}$ is equal to 1,
\beq
{\cal P}_*(\rL)\propto \delta(\rL-\rL^{(*)}).
\eeq
This is in a sharp contrast with Weinberg's conjecture (\ref{WC}).

There seems to be no reason to expect that the value of $\rL$ selected
by the condition (\ref{gammamax}) will fall into the anthropic range.
This approach is therefore unlikely to explain the smallness of $\rL$
or the cosmic time coincidence.

We also mention some alternative approaches.  Hawking \cite{Hawking}
suggested that the probability distribution for the observed values of
$\rL$ is given by Eq.(\ref{Pnucl}) with a minus sign in the
exponential and with $S(\rL)=-3/8\rL$, corresponding to a de Sitter
instanton of energy density $\rL$,
\beq
{\cal P}\propto \exp (3/8\rL).
\label{Haw}
\eeq
Since the Lorentzian continuation of this instanton describes an empty
universe dominated by the cosmological constant, it cannot be used to
describe the nucleation of the universe, so Eq.(\ref{Haw}) is
hard to justify.

Coleman \cite{Coleman} 
suggested that the Euclidean path integral of quantum gravity
is dominated by the lowest-energy de Sitter instantons connected by
Planck-size wormholes.  The resulting probability distribution is
\beq
{\cal P}\propto \exp\left[\exp (3/8\rL)\right].
\label{Col}
\eeq
Both expressions (\ref{Haw}),(\ref{Col}) have non-integrable peaks at
$\rL=0$ and thus do not satisfy Weinberg's conjecture.

\section{Quintessence}

We finally comment on models of quintessence with ``tracking'' solutions
which are now being extensively discussed in the literature 
\cite{Quintessence}.  
These models require a scalar field $Q$ with a potential $V(Q)$
approaching zero at large values of $Q$.
Note that this assumes that the cosmological constant problem 
has been solved by some mechanism, so that the true 
cosmological constant is set equal to 
zero (as in the case of power-law potentials
discussed in Section III).
A popular example of quintessence is an
inverse power-law potential of the form
\beq
V(Q)=\lambda M^{4+\beta}Q^{-\beta}
\eeq
with a constant $M\ll m_{pl}$.
The quintessence field $Q$ approaches an attractor ``tracking''
solution and evolves towards larger values on a cosmological timescale
$t$.  When $Q$ becomes comparable to $m_{pl}$, 
the universe gets dominated by $V(Q)$, and the
parameters of the model can be adjusted so that this happens at the
present epoch. 

It has been argued \cite{Zlatev} that quintessence models
do not suffer from the cosmic time coincidence problem, because the
time $t_Q$ of $Q$-domination is not sensitive to the initial
conditions.  This time, however, does depend on the details of the
potential $V(Q)$, and observers should be surprised to find themselves
living at the epoch when quintessence is about to dominate.  
More satisfactory would be a model in which the potential depends on
two fields, say $Q$ and $\phi$, with $\phi$ slowly varying in space,
making the time of $Q$-domination position-dependent.  We could
choose, for example,
\beq
V(Q,\phi)=\lambda M^{4+\beta -n}\phi^n Q^{-\beta}
\label{VQphi}
\eeq
for $Q\gg M$ and $V(Q,\phi)\sim\lambda M^{4-n}\phi^n$ for $Q\lesssim M$.

For this model to work, the initial conditions for the fields $\phi$
and $Q$ at the end of inflation should be different:  $\phi$ should be
spread over a range $\Delta\phi\gg m_{pl}$ as before, while $Q$ should
be concentrated at small values, $Q\ll m_{pl}$, so that it can get to
the tracking solution.  This can be arranged if $Q$ has a non-minimal
coupling to the curvature, ${1\over{2}}\xi RQ^2$.  Then $Q$ acquires
an effective mass $m_Q^2= 12\xi H^2$ during inflation, and its values
immediately after inflation are concentrated in the range $Q^2\lesssim
H^2/\xi$ (bounds on on the time variation of the gravitational
constant at late times require $\xi\lesssim 10^{-2}$ 
\cite{ch}).  The field $\phi$ is assumed to be
minimally coupled to the curvature ($\xi_\phi=0$), and its values are
randomized by quantum fluctuations during inflation.  This results in
a flat distribution (\ref{flatphi}), provided that $\lambda$ and $M$ are
sufficiently small.  

With these assumptions, a typical region of the universe after
inflation will have $Q\ll m_{pl}$ and $\phi\gg m_{pl}$.  In all such
regions, $\phi$ will remain nearly constant, while $Q$ will evolve
along the tracking solution, until the potential (\ref{VQphi})
dominates the universe.  This happens at $Q\sim m_{pl}$.  The energy
density at the time of $Q$-domination is 
\beq
\rho_Q\sim\lambda M^{4+\beta-n}m_{pl}^{-\beta}\phi^n \propto \phi^n,
\eeq
and the {\it a priori} distribution for $\rho_Q$ is 
\beq
d{\cal P}_*(\rho_Q)\propto \rho_Q^{-{{n-1}\over{n}}} d\rho_Q,
\label{P*Q}
\eeq
as in (\ref{P*}).  The full distribution can be obtained as before
using Eq.(\ref{dP}).  Note, however, that the expression (\ref{nu})
for $\nu(\rho_Q)$ cannot be used in this case, because the evolution
of density perturbations is different in models with an evolving
$\rho_Q$ and with $\rL=const$.  Press-Schechter formalism has been
applied to structure formation in quintessence models by Wang and
Steinhardt \cite{Wang}, and their results can be easily adapted to the
calculation of $\nu(\rho_Q)$ in a specific quintessence model.
The cutoff of the growth of density perturbations at $t\sim t_Q$ in
quintessence models is milder than that in models with a constant
vacuum energy density, and we expect the peak of the probabilty
distribution for $\rho_Q$ 
to be shifted dowards larger values.  
The qualitative character of the
distribution is expected to be unchanged, and in particular the cosmic
time coincidence (\ref{tGtL}) is likely to hold for a wide range of
model parameters.

\section{Conclusions}

The results of our analysis include some bad news and some good news.
The bad news is that Weinberg's conjecture (a flat {\it a
priori} probability distribution ${\cal P}_*(\rL$)) is not generally valid.
This conclusion applies both to models with slowly varying potentials
and to models with an ensemble of disconnected universes having
different (constant) values of $\rL$.  We regard this as bad news
because, without Weinberg's conjecture, the anthropic approach becomes
less predictive.  

In the quantum-cosmological approach, ${\cal P}_*(\rL)$ tends to select a
single value of $\rL$.  One can hope that this approach may provide an
explanation for a vanishing true cosmological constant, but one would
still have to find another mechanism to
explain a small but nonzero effective cosmological constant.
In the case of a slowly varying potential $V(\phi)$, the 
{\it a priori} distribution ${\cal P}_*$ depends on the shape of
the potential, which is of course highly uncertain.   (There is,
however, a wide class of potentials for which the conjecture does
apply.) 

The good news is that the cosmic time coincidence (\ref{tGtL}) is not
very sensitive to the shape of $V(\phi)$.  For a power-law potential,
$V(\phi)\propto \phi^n$, one finds that the probability distribution
for $t_G/t_\L$ is peaked at $t_G/t_\L\sim 1$ in the wide range
$1\lesssim n \lesssim 10$.  Moreover, for values of $n$ in the range
$2\lesssim n \lesssim 5$, the peak of the probability distribution
for $\rL$ is closer to the observationally suggested values than it is
for $n=1$ (corresponding to the Weinberg's conjecture).

We have also suggested an extension of quintessence models in which
the time of quintessence domination is determined by a slowly varying
scalar field.  The above conclusions apply to this class of models,
with minor modifications.

A common objection to anthropic arguments is that they are not testable.
It is therefore worth pointing out that models with with a scalar field
potential playing the role of the cosmological constant are falsifiable,
at least in principle.  Such models predict the existence of a
nearly massless, minimally coupled scalar field.  Fluctuations of this
field are produced during inflation with the same spectrum as gravitons
(and with half the energy density). Thus, for instance, if the energy
density in gravity waves is found to be in the range marginally allowed by 
nucleosynthesis (as it may happen in some models of quintessential
inflation \cite{pevi}), the existence of a massless field would be ruled
out; and with it the anthropic explanation for the time coincidence.

\section{Acknowlegements}

We are grateful to Tom Banks Pierre Binetruy 
and Steven Weinberg for useful discussions.
This work was supported by CIRIT under grant 1998BEAI400244,
(J.G.) and by the National Science Foundation (A.V.).


\begin{thebibliography}{99}
\bibitem{Foot0}
A similar argument for the coincidence of the present time $t_0$ with
the curvature domination time $t_\Omega$ in an open universe was given
in \cite{Tama}.
\bibitem{Foot2}
The function $\gamma(\rL)$ can be found as an eigenvalue of the
Fokker-Planck operator, as discussed in Refs. \cite{Starobinsky,LLM}.  With a
fixed-time cutoff, as in Eq.(\ref{Vt}), this function depends on
the choice of the time variable $t$ \cite{LLM,VW}.  This gauge dependence
is usually regarded as a problem and some attempts have been made to 
define gauge-independent probabilities in the case of a single eternally
inflating universe \cite{AV96}. It appears, however, that in the approach 
we adopted here for an ensemble of universes, it would be unreasonable to
require gauge-independence. The quantity $\gamma(\rho_Lambda)$ 
characterizes the rate of growth of the thermalized volume. Clearly,
it depends on the time variable with respect to which the rate is calculated.
The most natural choice appears to be the proper time which we have used 
throughout the paper. It is conceivable, however, that there may be other
consistent and reasonable approcaches which may give different answers for 
the probability distribution in the case of disconnected
universes. Investigation of this issue is left for further research.
\end{thebibliography}
\end{document}